\documentclass[entropy,article,submit,moreauthors,pdftex]{mdpi} 
\usepackage{amsmath}
\usepackage{latexsym}
\usepackage{amssymb}

\firstpage{1} 
\pubvolume{1}
\issuenum{1}
\articlenumber{0}
\pubyear{2021}
\copyrightyear{2021}
\preto{\abstractkeywords}{\nolinenumbers}
\history{Received: date; Accepted: date; Published: date}

\Title{Stability and evolution of synonyms and homonyms in signaling game}

\Author{Dorota Lipowska $^{1,\dagger,*}$\orcidA{} and Adam Lipowski $^{2,\dagger}$\orcidB{}}

\AuthorNames{Adam Lipowski, Dorota Lipowska}

\address{%
$^{1}$ \quad Faculty of Modern Languages and Literature, Adam Mickiewicz University in Pozna\'n, Poland; lipowska@amu.edu.pl\\
$^{2}$ \quad Faculty of Physics, Adam Mickiewicz University in Pozna\'n, Poland; lipowski@amu.edu.pl
}

\corres{Correspondence: lipowska@amu.edu.pl;}

\firstnote{These authors contributed equally to this work.}
\abstract{
Synonyms and homonyms appear in all natural languages. We analyse their evolution within the framework of the signaling game. Agents in our model use reinforcement learning, where probabilities of selection of a communicated word or of its interpretation depend on weights equal to the number of accumulated successful communications. 
When the probabilities increase linearly with weights, synonyms appear to be very stable and homonyms decline relatively fast. Such behaviour seems to be at odds with linguistic observations. A better agreement is obtained when  probabilities increase faster than linearly with weights.  Our results may suggest that a certain positive feedback, the so-called Metcalfe's Law,  possibly drives some linguistic processes.
Evolution of synonyms and homonyms in our model can be approximately described using a certain nonlinear urn model.
}

\begin{document}
\maketitle
\section{Introduction} 
Evolution and structure of language is often analysed using computational modelling \cite{cangelosi_2002,nolfi_2010,ulizia_2020}. A particularly appealing research paradigm is inspired by the idea that language might have spontaneously appeared in a population of communicating individuals, possibly with some adaptive features \cite{pinker_1990}. This standpoint prompted numerous analysis of multi-agent models, which mimic such  communication and try to infer the properties of the emerging language and its possible further evolution  \cite{ steels_2012experiments, gong2014modelling, kirby2014iterated}. 

In certain models of this kind, language emergence and evolution is studied using the signaling game \cite{lewis2002convention}, where  communicating agents must decide which signal (i.e., a word) to send or how to interpret the signal they have received. To cope with this, agents  very often use some form of the reinforcement learning  \cite{skyrms2010signals, lenaerts2005evolutionary, barrett2006numerical, franke2016evolution, muhlenbernd2012simulating, liplipplos, vaneecke2020reconceptualising}. 
Language that emerges in such models may provide a unique form-meaning mapping (in a signaling game terminology, it is a signaling system), but there are also other possibilities. In some cases, synonyms or homonyms can emerge,  destroying thus  the unambiguity of the form-meaning mapping. Neglecting some linguistic nuances \cite{ravin2000polysemy}, synonymy means that a single concept can be expressed by different words while in the case of homonymy, one word carries different meanings. 

Of course, synonyms and homonyms should not be considered undesirable or unrealistic as they exist in virtually every natural language, and are actually quite common.
For example, out of approximately 60,000 entries in Webster's Seventh New Collegiate Dictionary, about 35\% are homonyms with two or more meanings \cite{byrd1987tools}. The thesaurus based on Corpus of Contemporary American English (containing 385 million words) points to 30,000 words with a total of 370,000 synonyms \cite{davies2009385}. Let us notice, however,  
that in most cases synonymous words correspond to similar, not the same, meanings \cite{jackson2007words}.

It seems quite plausible that synonyms or homonyms change in time. For example, the frequency of their usage may change and gradually one form will be preferred over the other, and the latter eventually can even disappear. Although linguistic data are difficult to interpret, there are some indications that in natural languages synonyms are quite rare in contrast to homonyms, which appear to be more common \cite{hurford_2003, clark1990pragmatics}.  Some linguists even insist that true synonyms do not exist or at best are very rare compared to homonyms \cite{lyons1981language, goldberg1995constructions}. 
There are some arguments that the difference in the frequency of synonyms and homonyms may be due to evolutionary pressures favouring speakers rather than hearers \cite{hurford_2003}, or to language acquisition in  childhood \cite{markman1989categorization}. 

Let us notice that synonyms actually compete in a quite different way from homonyms, which can be demonstrated  already within the framework of the signaling game. While synonymous words compete for being selected by a speaker, for a homonymous word, it is the hearer's role to assign an appropriate interpretation. It is thus possible that such a difference can affect an overall dynamics of synonyms and homonyms and eventually result in different degrees of their prevalence. 

In the present paper, we examine evolution of synonyms and homonyms within the framework of the signaling game.  We show that this evolution depends on the selection mechanism that is implemented in signaling game.
Our results suggest that to be consistent with  linguistic observations, the dynamics of our model should implement a certain positive feedback, known in some marketing or economic contexts as Metcalfe's Law \cite{shapiro1998information}. Qualitatively, some of our numerical results can be better understood by referring to certain urn models.

\section {Model}
First, we briefly describe a certain urn model that will help us to understand some aspects of our mult-agent signaling game.
\subsection{Nonlinear urn model}\label{nonum}
Urn models were introduced by P\'olya \cite{eggenberger1923uber} and intensively studied since then \cite{pemantle2007survey}. In such models, one considers an urn with white and black balls.  At each step, one of the balls is drawn randomly from the urn and its colour observed. It is then placed back in the urn together with an additional ball of the same colour,  and the process is repeated. In the simplest version, the probability to select a ball of a given colour is proportional to the number of such balls in the urn. Particularly interesting for us, however, is  a nonlinear version of such a model,  where the selection probability is proportional to the number of balls raised to a certain power~$\alpha$ \cite{drinea2002balls}.
The case $\alpha>1$ can be interpreted as a positive feedback, commonly referred to as Metcalfe's law \cite{shapiro1998information}. In this case, the urn becomes dominated by balls of one colour. For $\alpha<1$, one might say that there is a negative feedback and the urn tends to the state with an equal number of balls of each colour. The orignal P\'olya urn model is equivalent to the $\alpha=1$ case  and it separates these two different regimes. 

To develop a heuristic understanding of the behaviour of such a nonlinear urn model \cite{drinea2002balls}, first we denote the number of white and black balls by $x(t)$ and $y(t)$, where $t$ is the total number of balls in the system.  The probability that the ball selected at time $t+1$ is white equals $\frac{x(t)^{\alpha}}{x(t)^{\alpha}+y(t)^{\alpha}}$. Thus, the expected change of $x(t)$ might be written as
\begin{equation} 
\Delta x(t)=E[x(t+1)-x(t)]=\frac{x(t)^{\alpha}}{x(t)^{\alpha}+y(t)^{\alpha}},
\label{e-drinea}
\end{equation}
and similarly for $y(t)$. Using the heuristic approximation $\frac{\Delta y(t)}{\Delta x(t)}=\frac{dy}{dx}$ we obtain

\begin{equation} 
\frac{dy}{dx}=\frac{y^{\alpha}}{x^{\alpha}}.
\label{e-drinea1}
\end{equation}
When $\alpha=1$, the solution of Eq.~(\ref{e-drinea1}) is $y=cx$, where $c$ is a certain constant. Integrating Eq.~(\ref{e-drinea1}) for $\alpha<1$, one obtains that $y/x$ goes to 1 in the long run. For $\alpha>1$, one obtains that $y/x$ goes to 0 or infinity \cite{drinea2002balls}.  As we will suggest in the following, such a nonlinear urn model can help us understand the stability and evolution of synonyms and homonyms, at least within the framework of the signaling game.
%%%%%%%%%%%%%%%%%%%%%%%%%%%%%%
%%%%%%%%%%%%%%%%%%%%%%%%%%%%%%

\subsection{Multi-agent signaling game with reinforcement learning}
In our model, we have a population of $N$~agents, which  play a variant of the signaling game trying to establish names for $n_o$~objects. Agents are placed at the sites of a network, and in the present paper we assume that it is a complete graph, where each site is connected, and thus can communicate, with all the remaining sites. Each agent~($A$) for each object~($o$) has an inventory, where it stores $n_w$~words ($i$) with their corresponding weights:
\begin{equation}
w(A)_{i,o}    \ \ \   A=1,\ldots, N \ \  i=1,\ldots, n_w \ \ o=1,\ldots, n_o
\label{weight}
\end{equation} 
In an elementary step, a randomly selected agent (the speaker) communicates with one of its neighbours (the hearer). The speaker ($S$) chooses an object ($o$) and from a corresponding inventory selects a word ($i$), with a probability of selection ($p(S)_{i,o}$)  proportional to the weight of this word ($w(S)_{i,o}$) raised to a certain power $\alpha$.  More precisely,
\begin{equation}
p(S)_{i,o}=\frac{w(S)^{\alpha}_{i,o}}{\sum^{n_w}_{j=1}w(S)^{\alpha}_{j,o}}
\label{selctprob}
\end{equation}
To interpret the communicated word~($i$), i.e., to select an appropriate object~($o$), the hearer~($H$)  takes into account the weights of the communicated word in all inventories and a similar form of the probability of selection is used:
\begin{equation}
p(H)_{i,o}=\frac{w(H)^{\alpha}_{i,o}}{\sum^{n_o}_{q=1}w(H)^{\alpha}_{i,q}}
\label{selctprob1}
\end{equation}

When hearer's interpretation (i.e., the object it selects) agrees with speaker's choice, the reward for their communicative success is that they both increase by unity the weights of the communicated word in their respective inventories, which improves the chances of choosing successful words in future communication attempts (reinforcement learning). To avoid an excessive increase of weights, which could result in stable configurations, we apply a population renewal, which means that   the selected agent with a certain (small) probability~$p$ does not become a speaker but instead is replaced with a newly created agent with all weights $w(A)_{i,o}=1$. The population renewal is not the only way to avoid a standstill as one can use some other techniques such as memory loss or lateral inhibition  \cite{spike2017minimal}.

To examine a time dependance in our model, a unit of time $t=1$ is defined as $N\cdot n_o$ of elementary steps, i.e., in a unit of time for each agent each object is on average  once selected by a speaker.

The model described above has already been used to examine the role of the network structure in the emergence of linguistic coherence \cite{liplipplos}. It was observed that $\alpha>1$  leads to a faster convergence than $\alpha=1$ but not necessarily to a linguistic coherence state (i.e., a signaling system).  A moderate intensity of a population renewal helps to  reach such a coherence, but too large  impedes the process. However, some other factors, such as the structure of the network of interactions or parameters $n_o$ and $n_w$, also affect the evolution of the model \cite{liplipplos}. We do not  examine our model for $\alpha<1$ because in this case the probabilities of selection increase sublinearly with accumulated weights, which inhibits reaching the linguistic coherence. As we  show in the following sections, parameter $\alpha$ qualitatively affects the evolution of synonyms and homonyms in our model. 

The nonlinear urn model described in the previous subsection plays only an auxiliary role. The signaling game model that we defined in this section  can be considered as a multi-urn generalization of such a single-urn model with weights $w(A)_{i,o}$ (Eq.~(\ref{weight})) corresponding to the number of balls.  Let us also note that the selection probabilites Eq.~(\ref{selctprob}) and  Eq.~(\ref{selctprob1}) are determined by the same set of weights. To select a word for an object, the speaker takes into account the weights of all words in the corresponding repository (the summation in Eq.~(\ref{selctprob})  is over words). On the other hand, to interpret the communicated word, the hearer takes into account the weights of this word in all repositories (the summation in Eq.~(\ref{selctprob1}) is over objects).   In some implementations of the signaling game, separate inventories are used for the selection of communicated words and for their interpretation \cite{muhl2014}.

%%%%%%%%%%%%%%%%%%%%%%%%%%%%%%
%%%%%%%%%%%%%%%%%%%%%%%%%%%%%%%
\section{Synonyms}
\begin{figure}
\includegraphics[clip, trim=0cm 5.8cm 11cm 1cm, width=\columnwidth]{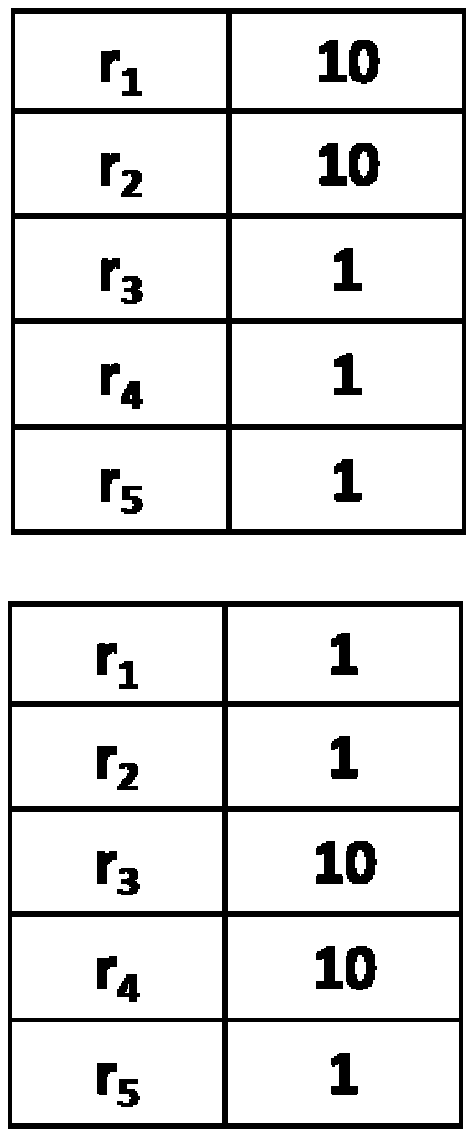}
\vspace{-0mm}
\caption{Two examplary agent's inventories of five words (left column) with weights (right column; $n_o=2$ and $n_w=5$). To examine the evolution of synonyms, we used  an initial configuration where in each agent's inventory, there is a pair of words with larger weights. When talking about object $O_1$ (or $O_2$), agents are initially inclined to communicate using synonymous words $r_1$ and $r_2$ (or $r_3$ and $r_4$).
}
\label{syno}
\end{figure}

To examine the evolution of synonyms, we prepared an initial configuration with synonyms, where for each object~($i$), agents have a set pair of words ($2i$ and $2i-1$) of weights larger than those of other words in the respective inventory (for simplicity, we chose two consecutive words for subsequent objects).  An example of such configuration  for $n_o=2$ and $n_w=5$ is shown in Fig.~\ref{syno}. Initially, thus,  agents are inclined to use one of those two words for an object. To monitor to what extent the synonymous structure persists in the system, we measured the parameter~$m$ defined as
\begin {equation}
m=\frac{1}{N n_o}\sum_{A=1}^N \sum_{i=1}^{n_o} \frac{|w(A)_{2i,i}-w(A)_{2i-1,i}|}{w(A)_{2i,i}+w(A)_{2i-1,i}}
\label{eq-s}
\end {equation}
This parameter measures the relative difference of weights of synonymous words. When synonyms persist in the system and have approximately equal weights, then $m$~is close to~0. When synonyms are eliminated,  typically one of the synonymous words loses importance (i.e., its weight) and $m$~increases toward unity.
%%%%%%%%%%%%%%%%%%%%%%%%%%%%%%%%%%
%%%%%%%%%%%%%%%%%%%%%%%%%%%%%%%

\begin{figure}
\includegraphics[width=\columnwidth]{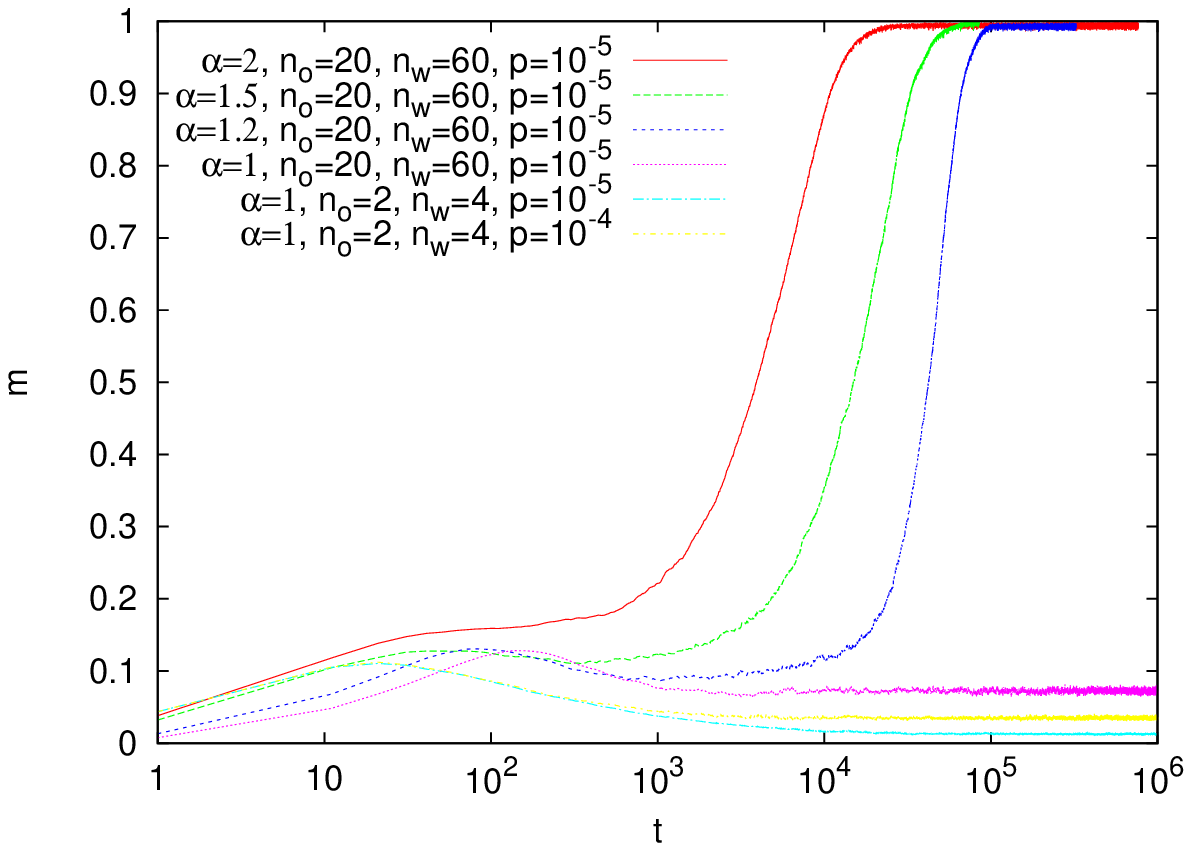}
\vspace{-0mm}
\caption{ Time evolution of synonyms as measured with the parameter~$m$ (Eq.~(\ref{eq-s})). Simulations were done for $N=10^3$ agents and the presented results are averages over 100 independent runs. Note the logarithmic scale on the horizontal axis.
}
\label{time-syno}
\end{figure}

Numerical simulations show that persistence of synonyms strongly depends on $\alpha$ (Fig.~\ref{time-syno}).   As simulations with $n_o=20$ and $n_w=60$ demonstrate, the characteristic life-time of synonyms increases for decreasing $\alpha$ and it seems to diverge for $\alpha\rightarrow 1$. For $\alpha=1$ synonyms seem to be very stable (even up to $t=10^6$) and such a behaviour was observed for several sets of parameters~$n_o$ and~$n_w$.

%%%%%%%%%%%%%%%%%%%%%%%%%%%%%%%%%%
%%%%%%%%%%%%%%%%%%%%%%%%%%%%%%%

Such a behaviour of synonyms can be understood referring to the nonlinear  urn model described in section \ref{nonum}. Assuming that for a given object the two competing synonyms have dominant weights, we can neglect other words in this inventory. Moreover, we can assume that these two words have  rather small weights in inventories corresponding to other objects. The weights of the first and second synonyms at time~$t$ are denoted by $x(t)$ and $y(t)$, respectively. The average change of $x(t)$ in one elementary step is equal to the probability that this  synonym will be selected by the speaker (which is $\frac{x(t)^{\alpha}}{x(t)^{\alpha}+y(t)^{\alpha}}$, because the two synonymous words dominate in the inventory) times the probability that this word will be correctly interpreted by the hearer (which equals~1, as we assumed that this word does not have large weights in any other inventory). Thus the average increase $\Delta x(t)$ has exactly the same form as in Eq.~(\ref{e-drinea}). Since the analogous equation can be written for $\Delta y(t)$, we obtain that the competition of synonyms in our model is equivalent to the nonlinear urn model (subject to some simplifying assumptions that we made).
From the analysis of the latter, it is thus clear that for $\alpha>1$, one of the synonyms gets eliminated while for $\alpha=1$, both of them are likely to persist.

\begin{figure}
\includegraphics[width=\columnwidth]{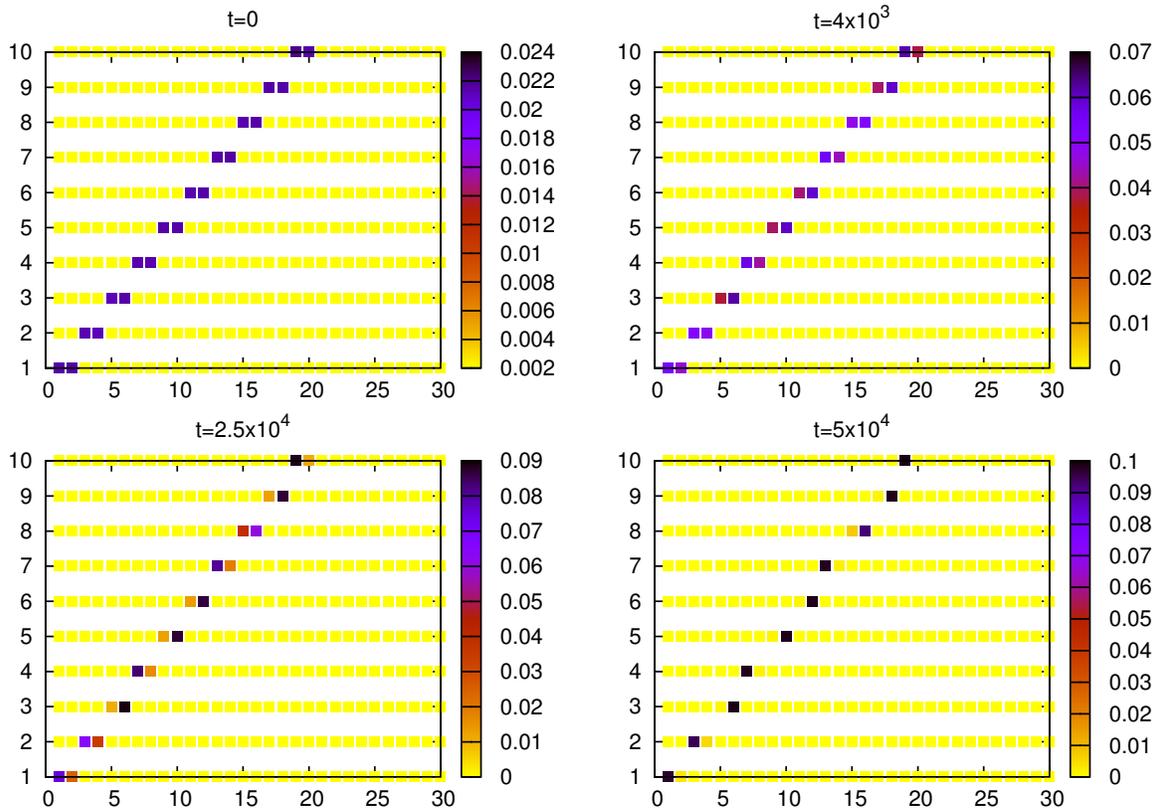}
\vspace{-0mm}
\caption{ Evolution of the weights (averaged over $N=10^3$ agents) for given objects and words. Simulations were done for $n_o=10$, $n_w=30$, $\alpha=1.5$, and $p=10^{-5}$. The initial configuration of weights includes 10 pairs of synonymous words. After $25\cdot 10^3$ steps, some asymmetry in weights can be noticed and at $t=5\cdot 10^4$, an almost perfect unique object-word mapping can be seen.
}
\label{syno-weights}
\end{figure}

\begin{figure}
\includegraphics[width=\columnwidth]{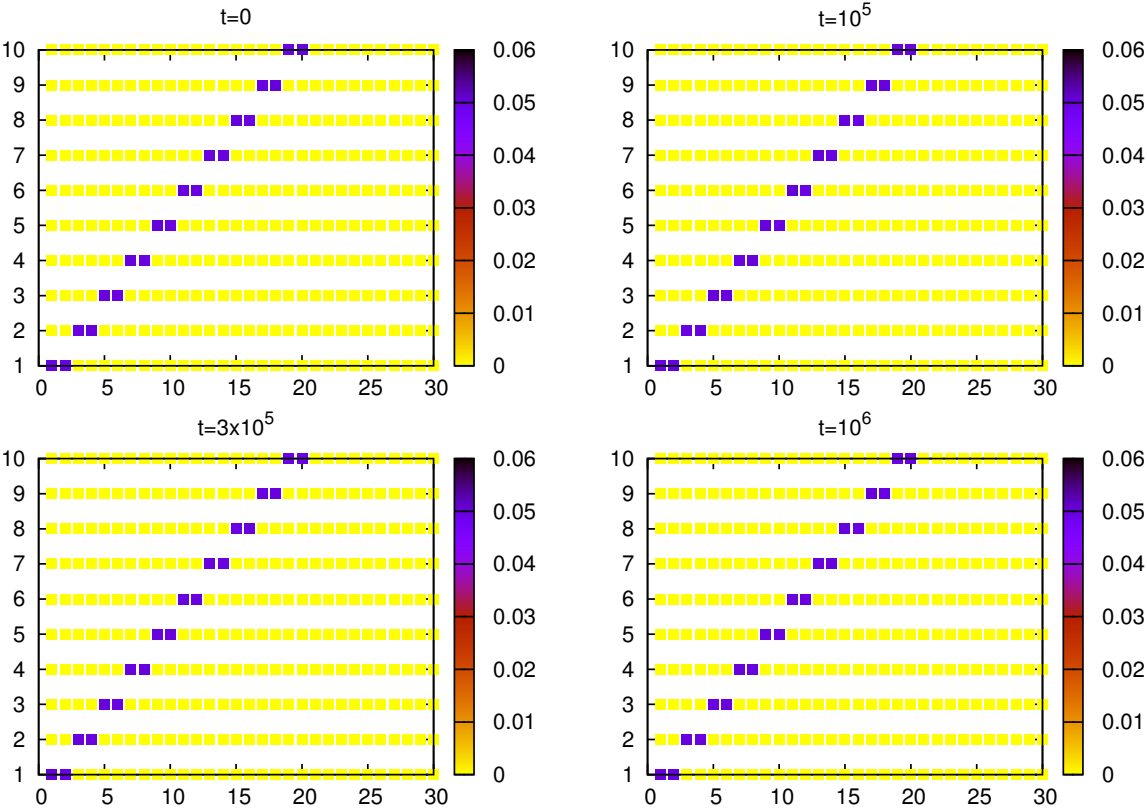}
\vspace{-0mm}
\caption{ Evolution of the weights (averaged over $N=10^3$ agents) for given objects and words. Simulations were done for $n_o=10$, $n_w=30$, $\alpha=1$, and $p=10^{-5}$. The initial configuration of weights includes 10 pairs of synonymous words. Evolution of the model shows that in this case synonyms are very stable.
}
\label{syno-weights1}
\end{figure}

To illustrate the evolution of the model with synonymous initial conditions, we plotted the weights of given objects and words (Fig.~\ref{syno-weights}, Fig.~\ref{syno-weights1}). The weights are normalized with the total weight (the sum of weights over all objects and words). Initally, for each object two words have their weights larger (than other words in the inventory) and mainly such synonyms are in use. For $\alpha=1.5$ (Fig.~\ref{syno-weights}), after approximately $25\cdot 10^3$ steps, a certain asymmetry is noticeable and gradually some of these synonymous words start to dominate. Further evolution leads to the formation of a unique object-word mapping (i.e, signaling system), where a single word is almost always used by a speaker to communicate a given object and this word is almost always correctly interpreted by a hearer. Since $\alpha=1.5$ is not much greater than unity, the process of elimination of synonyms is relatively slow.  
%%%%%%%%%%%%%%%%%%%%%%%%%%%%%%%%%%
%%%%%%%%%%%%%%%%%%%%%%%%%%%%%%%

Similar calculations for $\alpha=1$ (Fig.~\ref{syno-weights1}) show that in this case the pairs of synonyms exist basically unchanged for at least $t=10^6$ steps, in agreement with the above analysis based on the nonlinear urn model (Fig.~\ref{syno-weights1}).

%%%%%%%%%%%%%%%%%%%%%%%%%%%%%%%%%%
%%%%%%%%%%%%%%%%%%%%%%%%%%%%%%%

%%%%%%%%%%%%%%%%%%%%%%%%%%%%%%%

%%%%%%%%%%%%%%%%%%%%%%%%%%%%%%%
\section{Homonyms}
\begin{figure}
\includegraphics[clip, trim=0cm 10cm 12cm 3cm, width=\columnwidth]{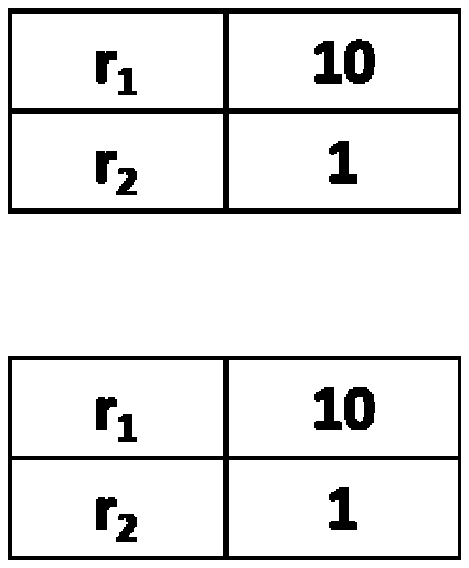}
\vspace{-0mm}
\caption{ Two examplary agent's inventories of two words (left column) with weights (right column; $n_o=2$ and $n_w=2$). To examine the evolution of homonyms, we used  an initial configuration where agent's inventories are grouped in pairs containing the same word with a weight larger than those of other words in these two inventories.  When talking about any of the objects $O_1$ or $O_2$, agents are initially inclined to communicate using the word~$r_1$. However, as a homonym, this word in both inventories has equal weights and the probability of its correct interpretation is only~0.5.
}
\label{homo}
\end{figure}

To examine the evolution of homonyms, we prepared the initial configuration with homonyms, i.e., this time one word of a large weight is associated with a chosen pair of objects. An example of such a configuration is shown in Fig.~\ref{homo}.

%%%%%%%%%%%%%%%%%%%%%%%%%%%%%%%%%%
\vspace{5mm}
%%%%%%%%%%%%%%%%%%%%%%%%%%%%%%%
In the case of homonyms, the communicated word can be incorrectly interpreted by the hearer, which reduces the success rate~$s$ of communicating agents, where $s$~is defined as a fraction of successfull communication attempts. Thus, such a parameter is  suitable for monitoring the stability of homonyms.

\begin{figure}
\includegraphics[width=\columnwidth]{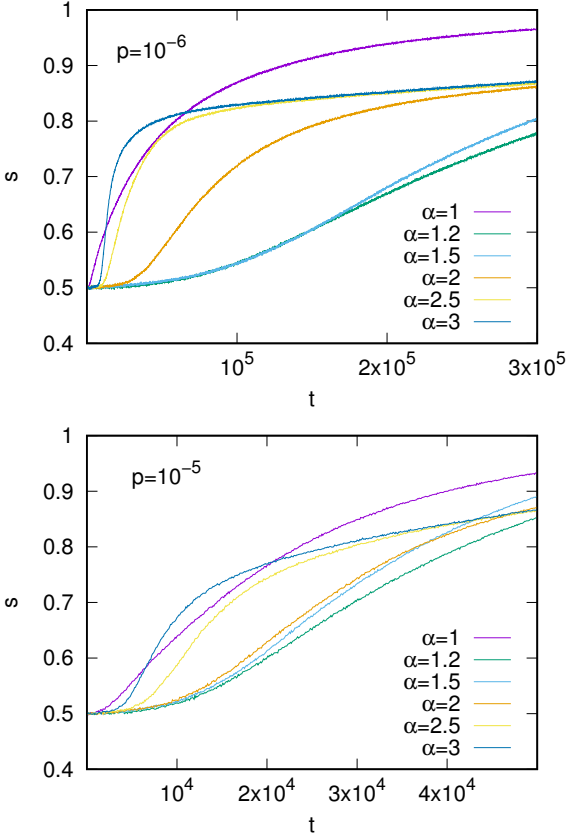}
\vspace{-0mm}
\caption{ Time evolution of the success rate~$s$ for the initial configuration with homonyms. Simulations were done for $N=10^3$ agents, and $n_o=2$, $n_w=2$. The presented results are averages over 100 independent runs.  
}
\label{time-homo}
\end{figure}

We made numerical simulations for the simplest possible case of $n_o=2$ and $n_w=2$ (a single homonym) but we believe that our conclusions concerning the stability of homonyms may also hold in other cases. We examined several values of $\alpha$ and the population renewal probability  $p=10^{-5}$ or $10^{-6}$ (Fig.~{\ref{time-homo}}). As expected, initially $s$~is close to 0.5  (see Fig.~\ref{homo}), however, it increases over simulation time, which indicates that homonymy is gradually eliminated and a unique object-word mapping is formed.

%%%%%%%%%%%%%%%%%%%%%%%%%%%%%%%%%%
%%%%%%%%%%%%%%%%%%%%%%%%%%%%%%%

Let us notice the nonmonotonic behaviour with respect to the parameter $\alpha$. For $\alpha=1$, the success rate~$s$ rapidly increases, but the increase is much smaller for $\alpha=1.2$ and 1.5. For $\alpha\geq 2$, the increase of~$s$ is again relatively fast. Since an increase in the success rate~$s$ is related to elimination of homonyms, our simulations show that homonyms are the most stable  for $\alpha=1.2$ and~$1.5$. 

\begin{figure}
\includegraphics[clip, trim=0cm 6cm 12cm 7cm,width=0.5\columnwidth]{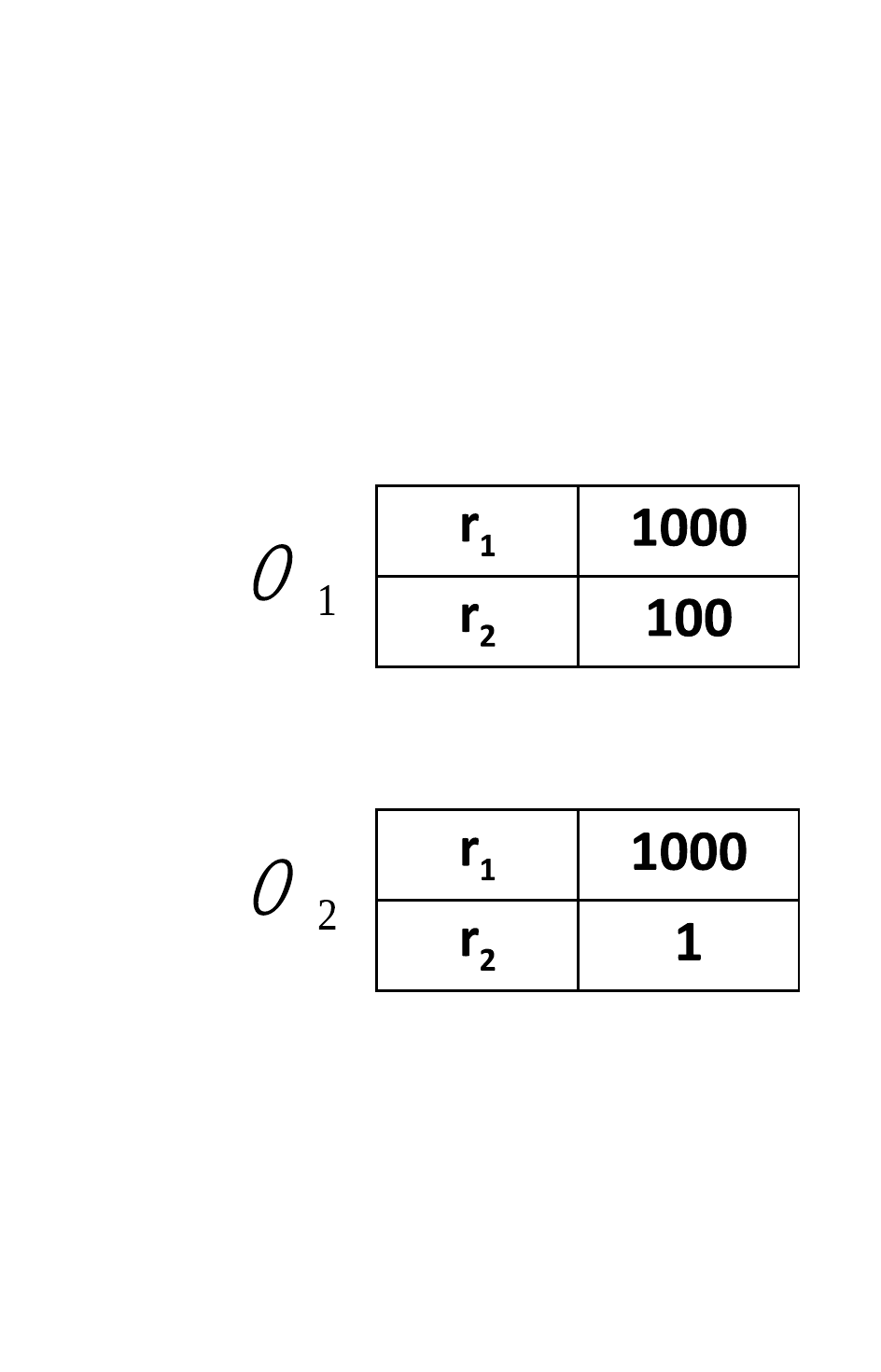}
\vspace{-0mm}
\caption{ The non-homonymous word~$r_2$ acquired a weight larger in the first inventory (for the object $O_1$) than in the second one (for the object $O_2$). When it happen to be communicated by a speaker who has chosen the object $O_1$, it is likely to be correctly interpreted by a hearer, which will result in a further increase in its weight in the first inventory.
}
\label{homo1}
\end{figure}

However, it is in our opinion somewhat surprising that for $\alpha=1$, we observe a fast increase of~$s$,  which is much in contrast with the stability of synonyms in this case. 
Some understanding of such behaviour can be inferred from the analysis of a certain urn model, which should approximately reflect the beginning of homonym elimination. The process starts when in a certain inventory a non-homonymous word acquires a sufficiently large weight and further evolution makes it a dominant word, which suppresses a homonym. For simplicity, let us consider only the inventory, where such a process takes place (first inventory in Fig.~\ref{homo1}).
Denoting the weights of the competing words by $x(t)$ (homonymous word) and $y(t)$ (non-homonymous word), we can write the following equations  that  describe their average change in a single step:
\begin{equation}
\label{homo-eq1}
\Delta x(t)=p_h\frac{x(t)^{\alpha}}{x(t)^{\alpha}+y(t)^{\alpha}}, \ 
\ \Delta y(t)=p_{nh}\frac{y(t)^{\alpha}}{x(t)^{\alpha}+y(t)^{\alpha}}
\end{equation}
The factors $p_h$ and $p_{nh}$ are the probabilities that homonymous and non-homonymous words, respectively,  will be correctly interpreted. 
The terms $\frac{x(t)^{\alpha}}{x(t)^{\alpha}+y(t)^{\alpha}}$ and $\frac{y(t)^{\alpha}}{x(t)^{\alpha}+y(t)^{\alpha}}$  in Eq.~(\ref{homo-eq1}) are the probabilites to select homonymous and non-homonymous words, respectively.  From the above equations and with the same heuristic replacement as in Eq.~(\ref{e-drinea1}), we obtain: 
\begin{equation}
\label{homo-eq2}
\frac{dx}{dy}=\frac{p_h}{p_{nh}}\frac{x^{\alpha}}{y^{\alpha}}
\end{equation}
%%%%%%%%%%%%%%%%%%%%%%%%%%%%%%%%%%
\vspace{5mm}
%%%%%%%%%%%%%%%%%%%%%%%%%%%%%%%
Let us notice that when a non-homonymous word that acquired a large weight in a certain inventory is communicated, it is very likely that it will be correctly interpreted (as illustrated in Fig.~\ref{homo1}). Thus we expect that $p_{nh}$ is close to~1 and $p_h$ is close to~0.5  because at the beginning of the homonym elimination, both homonyms are symmetric (i.e., of equal weights). 
Integrating Eq.~(\ref{homo-eq2}), for $\alpha=1$  we obtain  $x=C y^{p_h/p_{nh}}$, where $C$~is a certain constant.
Thus, for $t\rightarrow \infty$  and $p_h/p_{nh}<1$ (which is the case, as justified above), the ratio~$x/y$ equals~0. 
In other words, the weight of a non-homonymous word will dominate that of a homonymous word. It means that the non-homonymous word, even with a much smaller weight initially, will overtake a homonym in the long run.  

\begin{figure}
\includegraphics[width=\columnwidth]{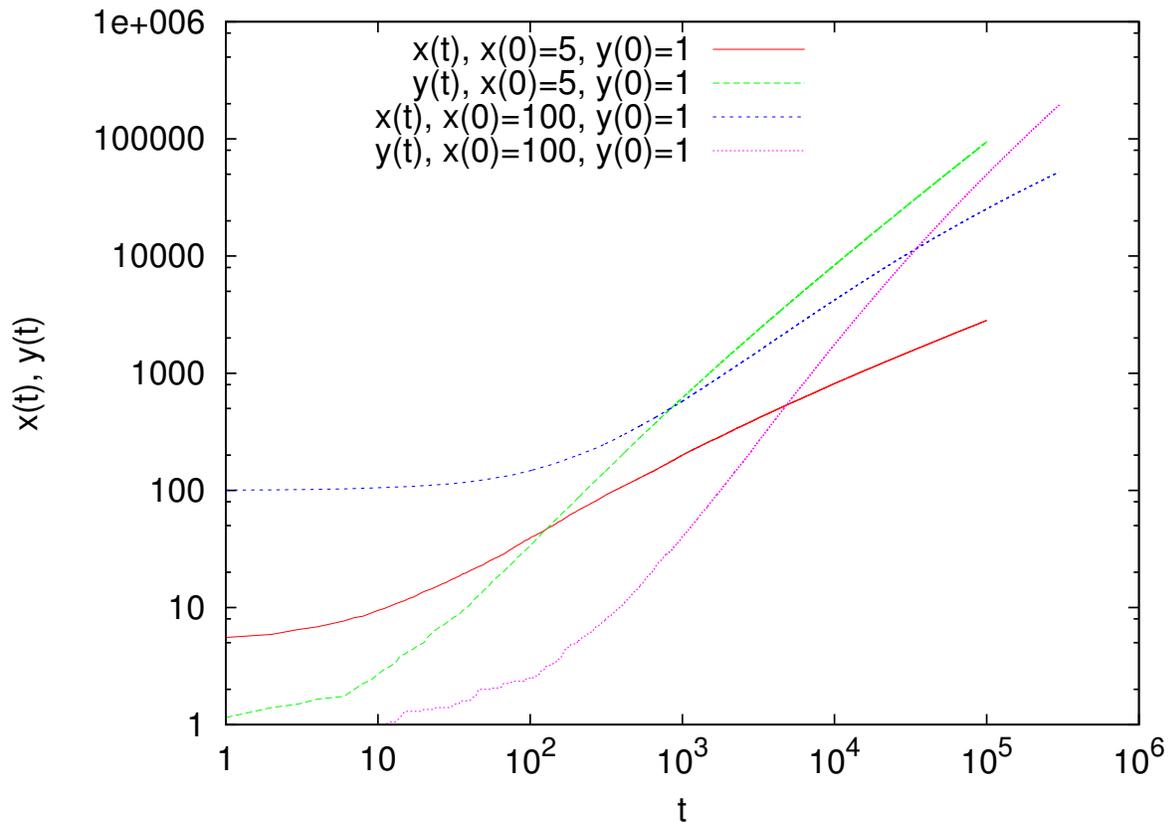}
\vspace{-0mm}
\caption{ Time evolution of the average weights $x(t)$ and $y(t)$ for the urn model with a stochastic addition (see the text).  The presented results are averages over 20 independent runs.
}
\label{weights}
\end{figure}

In our opinion, this result is by no means obvious and because the differential equation approach (Eq.~(\ref{homo-eq2})) is based on some heuristic assumptions, we made simulations of the urn model with the rules corresponding to Eq.~(\ref{homo-eq1}) with $\alpha=1$. Namely, there are $x(t)$ white balls and $y(t)$ black balls in the urn. We select randomly one of them. If it is a white one, then with probability~1/2 we add a white ball to the urn (and with probability~1/2 we do nothing).  If  the selected ball is black, we add one black ball to the urn. Numerical results for this urn model support the above analysis (Fig.~\ref{weights}). We can see that even when initial conditions strongly favour  $x(t)$ (homonymous word), it is   $y(t)$  (non-homonymous word) that dominates in the long run. 

%%%%%%%%%%%%%%%%%%%%%%%%%%%%%%%%%%
%%%%%%%%%%%%%%%%%%%%%%%%%%%%%%%

When $\alpha>1$, the prefactor $\frac{p_h}{p_{nh}}$ in Eq.~(\ref{homo-eq2}) is unimportant and similarly as for the nonlinear urn model Eqs.~(\ref{e-drinea}-\ref{e-drinea1}), we obtain that $y/x$ is equal to~0 or~$\infty$. However, in the context of the evolution of homonyms, we have the initial condition that favours $x(t)$ and thus for $\alpha>1$ one gets $y/x\rightarrow 0$ and homonyms appear to be stable against words that due to fluctuations acquired a relatively large weight. We would like to emphasize that this conclusion is based on the heuristic analysis of the nonlinear urn model.  The signaling game with its stochastic fluctuations and additional factors such as population renewal only to some extent agrees with this behaviour. Indeed, as demonstrated in Fig.~\ref{time-homo}, for $\alpha>1$ homonyms seem to be more stable than for $\alpha=1$, but in the long run, they are also  gradually eliminated.

\begin{figure}
\includegraphics[width=\columnwidth]{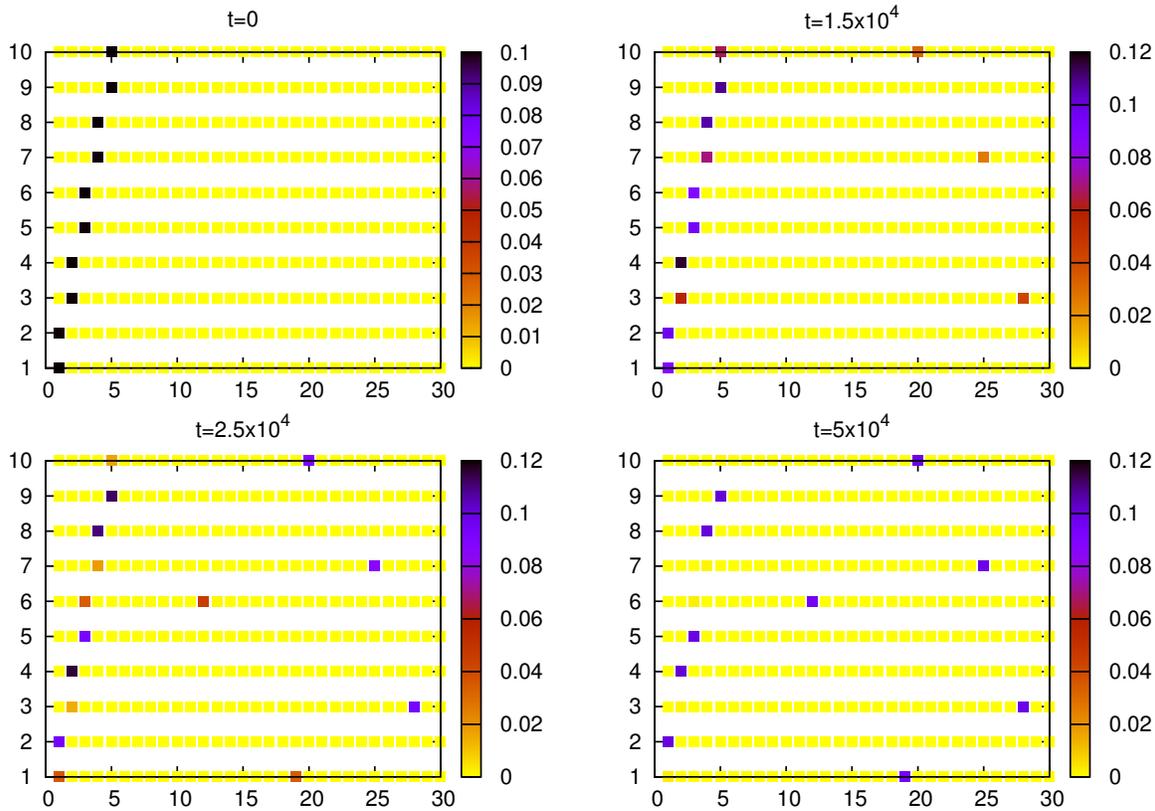}
\vspace{-5mm}
\caption{ Evolution of the weights (averaged over $N=10^3$ agents) for given objects and words. Simulations were done for $n_o=10$, $n_w=30$, $\alpha=1.5$, and $p=10^{-5}$ and the initial configuration of weights corresponds to 5~homonyms,  each with 2~meanings (objects). After $15\cdot 10^3$ steps, non-homonymous words for some objects acquire increasing weights and gradually dominate homonymous ones. Around $5\cdot 10^4$, we can see an almost perfectly unique object-word mapping.
}
\label{homo-weights}
\end{figure}

\begin{figure}
\includegraphics[width=\columnwidth]{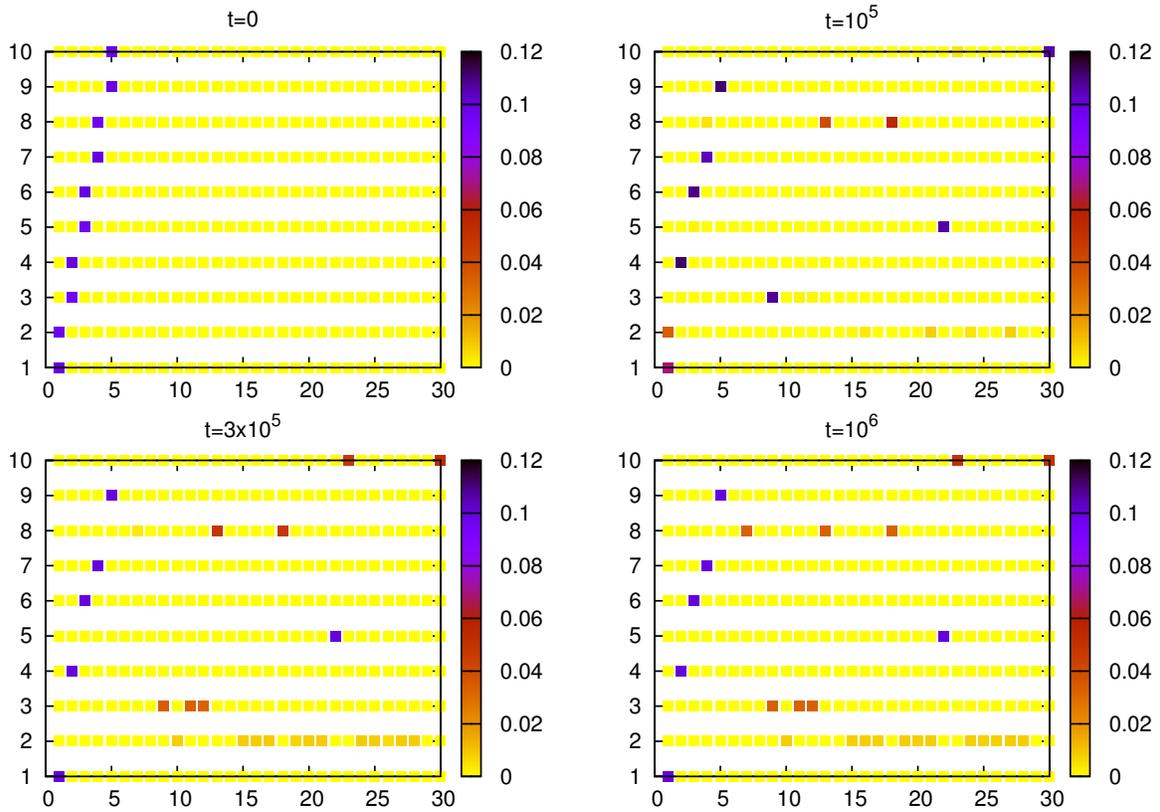}
\vspace{-5mm}
\caption{ Evolution of the weights (averaged over $N=10^3$ agents) for given objects and words. Simulations were done for $n_o=10$, $n_w=30$, $\alpha=1$ and $p=10^{-5}$ and the initial configuration of weights corresponds to 5~homonyms,  each with 2~meanings (objects). Around $3\cdot 10^5$, homonyms are basically eliminated but the emerging language contains synonyms (even multiple).
}
\label{homo-weights1}
\end{figure}

Finally, we present the evolution of weights in simulations  with a larger number of objects and words. 
The initial weights detrmine 5 homonyms (each representing two objects). For $\alpha=1.5$ (Fig.~\ref{homo-weights}) at $t=15\cdot 10^3$, we can notice that for some objects  non-homonymous words acquired substantial weights and they gradually suppress the homonym. Around $t=5\cdot 10^4$ all homonyms are eliminated and a unique object-word mapping is formed. However, in this case homonyms are relatively  long-lived (see Fig.~\ref{time-homo}).   We made similar calculations  for $\alpha=1$ (Fig.~\ref{homo-weights1}), in which case homonyms seem to be eliminated a bit faster. The resulting language contains also some synonyms, and because $\alpha=1$, they are likely to persist. 

%%%%%%%%%%%%%%%%%%%%%%%%%%%%%%%%%%
%%%%%%%%%%%%%%%%%%%%%%%%%%%%%%%

%%%%%%%%%%%%%%%%%%%%%%%%%%%%%%%%%%
%%%%%%%%%%%%%%%%%%%%%%%%%%%%%%%
Fig. \ref{syno-weights1}  and Fig. \ref{homo-weights1} show that the signaling game with $\alpha=1$ is characterized by very stable synonyms and a relatively fast elimination of homonyms, which seems to be different from linguisitc observations of natural  languages. With moderate nonlinearity (e.g., $\alpha=1.5$), both synonyms and homonyms decay on comparable time scales (Fig. \ref{syno-weights}, Fig. \ref{homo-weights}), which seems to be more adequate.
\vspace{5mm}

Finally, let us notice that the initial configurations that we used have maximal numbers of synonyms or homonyms and differ, of course, from existing languages, where the frequency of such structures is much smaller. Such a choice, however, simplifies computational analysis and enables us to monitor the evolution of syno- and homonyms. We expect that a signaling game with a more realistic initial configuration should exhibit an analogous behaviour but its computational analysis would be much more demanding, e.g., due to a much larger number of objects ($n_o$) that would need to be considered.

%%%%%%%%%%%%%%%%%%%%%%%%%%%%%%
%%%%%%%%%%%%%%%%%%%%%%%%%%%%%%%
\section{Conclusions}
In the present paper, we suggest that the presence of synonoyms and homonyms in natural languages may give us some valuable clues as to the nature of the mechanisms that drive linguistic processes.
Within the framework of the signaling game, we argue  that the reinforcement learning should operate in the super-linear regime (\mbox{$\alpha>1$}) with probabilities of selections increasing faster than linearly with the accumulated weights. The  linear regime  ($\alpha=1$) would, instead, lead to languages with very stable synonyms and relatively fast decaying homonyms, which is probably at odds with some linguistic observations \cite{hurford_2003}.

We have argued that certain aspects of the dynamics of synonyms and homonyms in the signaling game can be understood due to the similarity to the nonlinear urn model \cite{drinea2002balls}.  The relationship is not entirely straightforward: the monopolistic regime that appears in the nonlinear urn model translates as a removal of synonyms but for homonyms, it actually protects their stability.
The signaling game is of course more complex than the urn model and an analysis of the latter  provides only a limited insight into the evolution of synonyms and homonyms in the signaling game. Let us also note that our model contains several parameters such as $N$, $n_o$, $n_w$, $\alpha$, or $p$. Together with the choice of the network of interactions (it is a complete graph in our study) or the initial values of weights, these are all the factors influencing the evolution of the model. We did only very limited numerical analyses for the values of parameters that, in our opinion, correspond to some generic (within some bounds) behaviour of the model. Of course, more complete analyses of the model would be desirable.

As we have already mentioned, synonyms, contrary to homonyms, do not reduce the communicative  efficiency in the signaling game.  However, such a property should  not be considered as implying that synonyms are stable and homonyms are not. Although for $\alpha=1$, we indeed observe stable synonyms and unstable homonyms, the reverse situation takes place for $\alpha>1$ (in which case the impact of synonyms and homonyms on the communicative efficiency is the same).  As a possible further extension, one could consider the possibility that synonyms are less stable if there is a cost related to the size of the inventory used, i.e., when there is a pressure against using more words than necessary for perfect communication. Moreover, homonymy can be evolutionarily stable if the signaling game is modified, for example, when there are additional sources of information that  can be used for disambiguation by the hearer  \cite{o2015ambiguity,santana2014ambiguity,muhl2021}.

Perhaps an interesting question is why nonlinear ($\alpha>1$) rather than (maybe naively expected) linear ($\alpha=1$) feedback drives linguistic processes. In marketing or economic contexts, a competition between, e.g.,  video recording formats, operating systems, and even keyboard types  is often discussed, in which cases the value of the system grows probably faster than linearly with the number of users (Metcalfe's Law)   \cite{shapiro1998information, arthur1994increasing}. In the signaling game, it would mean that a benefit of using a certain word (and thus a probability of its future selection) increases faster than linearly with the number of successful communications. Considering   the complexities of language evolution, with its various social, biological, and cognitive aspects, it seems quite likely.  

%%%%%%%%%%%%%%%%%%%%%%%%%%%%%%%%%%
%%%%%%%%%%%%%%%%%%%%%%%%%%%%%%%%%%

%\bibliography{abbrev_titles,biblio}

%\input{bipartitioning.bbl}

%%%%%%%%%%%%%%%%%%%%%%%%%%%%%%%%%%
%%%%%%%%%%%%%%%%%%%%%%%%%%%%%%%%%%

%%%%%%%%%%%%%%%%%%%%%%%%%%%%%%%%%%%%%%%%%%
%%%%%%%%%%%%%%%%%%%%%%%%%%%%%%%%%%%%%%%%%%
\authorcontributions{conceptualization, D.L. ; methodology, A.L.; software, A.L. and D.L.; validation, A.L. and D.L.; investigation, A.L.  and D.L..; writing--review and editing, A.L. and D.L.; visualization, A.L. All authors have read and agreed to the published version of the manuscript.'', please turn to the  \href{http://img.mdpi.org/data/contributor-role-instruction.pdf}{CRediT taxonomy} for the term explanation. Authorship must be limited to those who have contributed substantially to the work reported.}

%%%%%%%%%%%%%%%%%%%%%%%%%%%%%%%%%%%%%%%%%%
\conflictsofinterest{The authors declare no conflict of interest.} 
%%%%%%%%%%%%%%%%%%%%%%%%%%%%%%%%%%%%%%%%%%
\reftitle{References}

%=====================================
% References, variant A: external bibliography
%=====================================
%\externalbibliography{yes}
%\nocite{*}
%\bibliographystyle{mdpi}
%\bibliography{abbrev_titles,biblio}

%=====================================
% References, variant B: internal bibliography
%=====================================
%\begin{thebibliography}{999}

%% Reference 1
%\bibitem[Author1(year)]{ref-journal}
%Author1, T. The title of the cited article. {\em Journal Abbreviation} {\bf 2008}, {\em 10}, 142--149.
%% Reference 2
%\bibitem[Author2(year)]{ref-book}
%Author2, L. The title of the cited contribution. In {\em The Book Title}; Editor1, F., Editor2, A., Eds.; Publishing House: City, Country, 2007; pp. 32--58.
%\end{thebibliography}

%% for journal Sci
%\reviewreports{\\
%Reviewer 1 comments and authors’ response\\
%Reviewer 2 comments and authors’ response\\
%Reviewer 3 comments and authors’ response
%}

%%%%%%%%%%%%%%%%%%%%%%%%%%%%%%%%%%%%%%%%%%
\end{document}